\documentclass[fleqn]{elsart}

\newcommand{\be}{\begin{equation}}
\newcommand{\ee}{\end{equation}}
\newcommand{\ba}{\begin{eqnarray}}
\newcommand{\ea}{\end{eqnarray}}
\def\bs{\begin{subequations}}
\def\es{\end{subequations}}
\def\a{\alpha}

\def\vp{\varphi}
\def\vr{\varrho}

\def\cE{{\cal E}}

\def\cL{{\cal L}}

\def\cM{{\cal M}}

\def\B{\Box}
\newcommand{\Eq}[1]{(\ref{#1})}

\def\rme{e}
\def\rmd{d}
\def\rmi{i}

\begin{document}

\begin{frontmatter}

\rightline{\small AEI-2010-006 \hfill arXiv:1012.1244}
\vspace{1cm}

\title{Gravity on a multifractal}

\author{Gianluca Calcagni},
\ead{calcagni@aei.mpg.de}
\address{Max Planck Institute for Gravitational Physics (Albert Einstein Institute),\\
Am M\"uhlenberg 1, D-14476 Golm, Germany}
\begin{abstract}
Despite their diversity, many of the most prominent candidate theories of quantum gravity share the property to be effectively lower-dimensional at small scales. In particular, dimension two plays a fundamental role in the finiteness of these models of Nature. Thus motivated, we entertain the idea that spacetime is a multifractal with integer dimension 4 at large scales, while it is two-dimensional in the ultraviolet. Consequences for particle physics, gravity and cosmology are discussed.
\end{abstract}


\end{frontmatter}

In 1884, Edwin A.\ Abbott published his satirical novella \emph{Flatland: A Romance of many dimensions}, where a Square living in the $(2+1)$-dimensional Flatland envisions different geometries. While it is easy for it to imagine worlds of lower dimensions such as Pointland and Lineland, it takes the intervention of a Sphere to have the Square realize the possibility of Spaceland (our world) and even more fantastic cosmos which even the Sphere cannot fathom. 

This book has been entertaining generations of teachers, mathematicians and physicists, keeping vivid in the public imagination the possibility that the universe, after all, might be more than a matter of spheres. In fact, the notion of higher dimensions has been considered most seriously by the scientific community, from Kaluza--Klein to brane-world scenarios. The latter can be motivated by perturbative string theory, where the number of spacetime dimensions is higher than four. The brane-world has been a popular playground where issues such as the hierarchy problem have found fresh insight \cite{P07_bw}.

On the other side of the story, models in lower dimensions are extremely helpful in addressing a number of physical and technical problems which are harder to tackle in $4D$. However, the dimensionality of spacetime is a fixed ingredient, so while in the case of brane and string scenarios the unobserved extra dimensions are explained via compactification or other mechanisms, lower-dimensional theories are typically regarded as toy, albeit very interesting, models of reality.

Nevertheless, there is another meaning in which a model can be ``lower-dimensional''. Independent theories such as causal dynamical triangulations, asymptotically safe gravity, spin-foam models, and Ho\v{r}ava--Lifshitz gravity all exhibit a running of the spectral dimension $d_{\rm S}$ of spacetime such that at short scales $d_{\rm S}\sim 2$ \cite{P07_spe}. This number is no chance and plays an important role in quantum gravity, not only in reference to the richness of worldsheet string theory, but also because gravity as a perturbative field theory is renormalizable near two dimensions \cite{P07_GKT}.

Is it possible to construct a field theory of matter and gravity which is effectively two-dimensional at small spacetime scales and four-dimensional in the infrared? Here we wish to argue for a positive answer, whose details can be found in \cite{P07_frac}. In homage to Abbott's novella, one would have liked to call the short-scale world a Lineland, but this would have been misleading. Nowadays we know that there exist geometric objects which are not curves or sheets or solids even if they have integer dimension. Fractals have required a revision and extension of the concept of ``dimension'', the Hausdorff definition being just one example. 
In many cases, often presented in rich pictorials, fractals have noninteger dimensions, but there exist instances where a dust or a curve can fill the ambient space enough to achieve integer dimensionality \cite{P07_Fal03}. Multifractals are objects with scale-dependent Hausdorff dimension.

The problem now is to encode in the structure of spacetime the dimensional flow typical of multifractals. This can be done by promoting the Lebesgue measure in the integral defining any field theory action to a generic Lebesgue--Stieltjes measure:
\be\label{P07_vr}
\rmd^D x\to \rmd\vr(x)\,,\qquad [\vr]=-D\a\neq -D\,,
\ee
where$\vr$ is a (possibly very irregular) distribution, square brackets denote the engineering dimension in momentum units, and $0<\a<1$ is a parameter which is related with the operational definition of the Hausdorff dimension $d_{\rm H}$ as follows. In fact, the latter determines the scaling of an Euclidean volume (or mass distribution) of characteristic size $R$, $V(R) \sim R^{d_{\rm H}}$. Taking $\vr\sim \rmd(r^{D\a})$,
\be
V(R)\sim\int_{D\textrm{-ball}}\rmd\vr_{\rm Eucl}(x)\sim \int_0^R \rmd r\, r^{D\a-1}\sim R^{D\a},
\ee
thus showing that 
\be
\a=\frac{d_{\rm H}}{D}.
\ee

Consider a Lorentz-covariant Lagrangian density $\cL$; this can be the total Lagrangian of gravity and matter on a manifold $\tilde\cM$ endowed with metric $g_{\mu\nu}$, where $\mu=0,1,\dots,D-1$ and $D$ is the topological (positive integer) dimension of $\tilde\cM$. To make the universe a multifractal $\cM$, we replace the standard measure in the action with a nontrivial Stieltjes measure:
\be
S=\int_\cM \rmd\vr(x)\,\sqrt{-g}\cL\,.
\ee
We assume $\cM$ has no boundary; the case with boundary should share most of the same qualitative features. If $\vr$ is absolutely continuous, it can be written as $\rmd\vr(x)=v(x)\rmd^Dx$, where $v$ is a Lorentz scalar. We can choose
\be\label{P07_mea}
v(X)= X^{D(\a-1)}+M^{D(1-\a)}\,,
\ee
where $M$ is a constant mass and $X=t$ or $X=|{\bf x}|$ depending on whether we want to define a ``timelike'' or ``spacelike'' multifractal. The metric $g_{\mu\nu}$ and the scalar $v$ are independent degrees of freedom which constitute the composite geometric structure (metric and fractal) of $\cM$. A fractal must shortly evolve to a smooth configuration. We expect $M$ to be about the Planck mass, although the lower bound from particle physics actually seems to be much lower, $M> 300\div 400~ {\rm GeV}$ \cite{P07_She09}.

Equation \Eq{P07_mea} is inspired by results in classical mechanics, according to which integrals on fractals can be approximated by Weyl or fractional integrals which, in turn, are particular Lebesgue--Stieltjes integrals. The order of the fractional integral $D\a$ has a natural interpretation in terms of the Hausdorff dimension of $\cM$ \cite{P07_1dfr}.
Fractional integrals find applications in a range of disciplines, from statistics to finance to engineering. In one dimension, different values of $\a$ mediate between full-memory ($\a=1$) and Markov processes ($\a=0$), where $\a$ corresponds to the fraction of states preserved at a given time. Loosely speaking, in our case it is the ``fraction of spacetime dimensionality'' felt by an observer living in $\cM$, which is equally divided among the $D$ directions for the isotropic weight \Eq{P07_mea}.

The Lorentz scalar $v$ may contribute a kinetic term if interpreted as part of the field dynamics, otherwise it is excluded from the calculus of variations. We must stress that Eq.~\Eq{P07_mea} is a very special case of Stieltjes measure and it is quite possible that realistic models with fractal behavior do not admit an absolutely continuous measure. In that case, it is not yet clear how to work out the details of the theory. 

Otherwise, properties of the class of models satisfying Eq.~\Eq{P07_mea} are well illustrated by a scalar field theory \cite{P07_frac}. The engineering dimension of the scalar field is zero when $\a$ has the critical value $\a=\a_*\equiv 2/D$. The dimension of spacetime is well constrained to be 4 from particle physics to cosmological scales and starting at least from the last scattering era. Therefore, $D=4$ for phenomenological reasons. The properties of the field causal propagator in configuration space depend on the value of $\a$. By making use of momentum-space results, one can see that the superficial degree of divergence of the Feynman-like diagrams of the theory is lower than in $4D$. This is promising but not sufficient to demonstrate the effectiveness and viability of a renormalization group flow. At any rate, at the classical level the system does flow from a lower-dimensional configuration to a smooth $D$-dimensional one. This is clear from the definition \Eq{P07_mea} of the measure weight and its scaling properties when $\a<1$, as already discussed. Therefore, at least the phenomenological valence of the model is guaranteed.

If $M\sim m_{\rm Pl}$, it is likely that UV effects be important only during the very early universe. This is suggested also by a minisuperspace analysis of the model \cite{P07_frac}, indicating that UV cosmological solutions with zero intrinsic curvature do not exist unless one allows for exotic matter fields (or condensates) violating the null energy condition. On the other hand, at late times an imprint of the nontrivial short-scale geometry might survive as a running cosmological constant. The latter appears as a source term in the Noether conservation law for the Hamiltonian $H$: in Minkowski, the energy of the system is
\be
\cE(t) = H(t)+\Lambda(t)=H(t)+\int^t\rmd t\int\rmd{\bf x}\,\dot v\cL\,.\label{P07_Lam}
\ee
In general, the physical $D$-momentum dissipates, which might constitute a unitarity problem at the quantum level. In nonrelativistic fractal models this is a direct result of the nonautonomous character of the action; translation invariance is broken \emph{explicitly}. In our relativistic scenario, $v$ is a scalar with implicit coordinate dependence, a geometric factor defined for a $D\a$-dimensional physical world which enters the definition of Poisson brackets. This is a key difference with respect to scalar-tensor theories and results in a \emph{deformation of the Poincar\'e algebra}. In this precise sense, also relativistic fractals break translation invariance. 

However, the system also admits a conservative interpretation. One can also regard $v$ as an independent ``dilaton-like'' field rescaling the total Lagrangian density in the $D$-dimensional ambient spacetime. In that case, one can define the Poisson brackets as usual (no Stieltjes measure within) and show that Poincar\'e invariance is preserved. Dissipation occurs relatively between parts of a \emph{conservative} system. Quantization would follow through, although an UV observer would experience an effective probability flow through his world-fractal.

We have just said that translation invariance is not broken explicitly in relativistic fractal field theory. For instance, the use of a nontrivial measure weight might lead to the idea that translation invariance be violated by the expression for the propagator. This is not the case, as we show here in more detail than in \cite{P07_frac}. Consider a free scalar field with action
\be
S_0=-\frac12\int\rmd\vr(x)\,\phi(x)\, f(\B)\,\phi(x)\,,
\ee
where we keep the kinetic operator $f(\B)$ general. The free Lorentzian partition function $Z_0$ in the presence of a local source $J$ is
\be\label{P07_z0}
Z_0[J]\equiv \int [{\cal D}\phi]\,\rme^{\rmi[S_0+\int\rmd\vr(x)\, J(x)\phi(x)]}\equiv\int [{\cal D}\phi]\,\rme^{\rmi S_J}\,.
\ee
Using the definition of the $D$-dimensional Dirac distribution with nontrivial measure
\be 
\delta_\vr(k)=\frac{1}{(2\pi)^{D}}\int\rmd\vr(x)\,\rme^{-\rmi k\cdot x}
\ee
and the Fourier--Stieltjes transform of the field $\phi(x)=(2\pi)^{-D}\int\rmd\vr(k)\,\tilde\phi(k)\,\rme^{\rmi k\cdot x}$, we obtain
\ba
S_J &=& \frac12\int\rmd\vr(x)\int\frac{\rmd\vr(k_1)}{(2\pi)^{D}}\int\frac{\rmd\vr(k_2)}{(2\pi)^{D}}\,\rme^{\rmi(k_1+k_2)\cdot x}\left[-\tilde\phi(k_1)f(-k_2^2)\tilde\phi(k_2)\right.\nonumber\\
&&\qquad\qquad\left.+\tilde J(k_1)\tilde\phi(k_2)+\tilde J(k_2)\tilde\phi(k_1)\right]\nonumber\\
&=& \frac12\int\frac{\rmd\vr(-k)}{(2\pi)^{D}}\left[-\tilde\phi(-k)f(-k^2)\tilde\phi(k)+\tilde J(-k)\tilde\phi(k)+\tilde J(k)\tilde\phi(-k)\right]\nonumber\\
&=& \frac12\int\frac{\rmd\vr(-k)}{(2\pi)^{D}}\left[-\tilde\vp(-k)f(-k^2)\tilde\vp(k)+\frac{\tilde J(-k)\tilde J(k)}{f(-k^2)}\right]\,,
\ea
where
\be
\tilde\vp(k)\equiv \tilde\phi(k)-\frac{\tilde J(k)}{f(-k^2)}\,.
\ee
Modulo the measure, we have followed exactly the same steps as in ordinary quantum field theory. Equation \Eq{P07_z0} becomes
\ba
Z_0[J] &=& \left\{\int [{\cal D}\vp]\exp\left[-\frac{\rmi}{2}\int\frac{\rmd\vr(-k)}{(2\pi)^{D}}\tilde\vp(-k)f(-k^2)\tilde\vp(k)\right]\right\}\nonumber\\
&&\times\exp\left[\frac{\rmi}{2}\int\frac{\rmd\vr(-k)}{(2\pi)^{D}}\frac{\tilde J(-k)\tilde J(k)}{f(-k^2)}\right]\nonumber\\
&=& Z_0[0]\,\exp\left[\frac{\rmi}{2}\int\frac{\rmd\vr(-k)}{(2\pi)^{D}}\frac{\tilde J(-k)\tilde J(k)}{f(-k^2)}\right]\,.
\ea
The exponent can be written as
\be\nonumber
\int\frac{\rmd\vr(-k)}{(2\pi)^{D}}\frac{\tilde J(-k)\tilde J(k)}{f(-k^2)}=\int\frac{\rmd\vr(-k)}{(2\pi)^{D}}\int\rmd\vr(x)\int\rmd\vr(y)\,\rme^{\rmi k\cdot(x-y)}\frac{J(x)J(y)}{f(-k^2)}\,,
\ee
so that, if $\vr(-k)=\vr(k)$, the free partition function reads
\be
Z_0[J]= Z_0[0]\,\exp\left[\frac{\rmi}{2}\int\rmd\vr(x)\int\rmd\vr(y)\,J(x) G(x-y) J(y)\right]\,,
\ee
where
\be
G(x-y)=\frac{1}{(2\pi)^{D}}\int\rmd\vr(k)\,\frac{\rme^{\rmi k\cdot (x-y)}}{f(-k^2)}\,.
\ee
Therefore, we have recovered the usual definition of the propagator as the solution of the Green equation 
\be
f(\B)\,G(x-y)=\delta_\vr(x-y)\,.
\ee

Other details and other features of the scalar field on an effective (multi)fractal spacetime can be found in \cite{P07_frac}.
 These properties are shared also by the gravitational sector when the latter is switched on. There, one can see that the bare Newton's constant is dimensionless for $\a=\a_*$, thus suggesting renormalizability \cite{P07_frac}. Physical implications of the UV propagator, renormalization, and the hierarchy problem will require further attention.


\end{document}